\documentclass[]{JHEP3} 

\usepackage{epsfig,multicol}
\usepackage{graphicx}
\usepackage{dcolumn}
\usepackage{bm}

\newcommand{\rlc}{r_{\rm lc}}

\newcommand{\zcl}{z_{\rm c}}
\newcommand{\Tcl}{T_{\rm c}}
\newcommand{\vcl}{v_{\rm c}}
\newcommand{\Tlss}{T_{\rm D}}
\newcommand{\alphalss}{\alpha_{\rm D}}
\newcommand{\etalss}{\eta_{\rm D}}

\newcommand{\tabox}[1]{\makebox[1.7cm][c]{#1}}
\newcommand{\taboxl}[1]{\makebox[3.7cm][c]{#1}}

\title{CMB observations in LTB universes: Part II\\
--- the kSZ effect in an LTB universe ---
}

\author{Chul-Moon Yoo
\\
	Yukawa Institute for Theoretical Physics, Kyoto University
Kyoto 606-8502, Japan
\\
Asia Pacific Center for Theoretical
Physics, Pohang, Gyeongbuk 790-784, Republic of Korea
	E-mail: \email{yoo{}@{}yukawa.kyoto-u.ac.jp}}

\author{Ken-ichi Nakao
\\
Department of Mathematics and Physics,
Graduate School of Science, Osaka City University,
3-3-138 Sugimoto, Sumiyoshi, Osaka 558-8585, Japan
	E-mail: \email{knakao{}@{}sci.osaka-cu.ac.jp}}

\author{Misao Sasaki
\\
	Yukawa Institute for Theoretical Physics, Kyoto University
Kyoto 606-8502, Japan
\\
Korea Institute for Advanced Study
207-43 Cheongnyangni 2-dong, Dongdaemun-gu, 
Seoul 130-722, Republic of Korea
	E-mail: \email{misao{}@{}yukawa.kyoto-u.ac.jp}}

\preprint{YITP-10-67,APCTP-Pre2010-005,OCU-PHYS-336,AP-GR-81}

\abstract{We study the kinematic Sunyaev-Zel'dovich (kSZ) effect in a 
Lem\^itre-Tolman-Bondi (LTB) universe model 
whose distance-redshift relation agrees with 
that of the concordance $\Lambda$CDM model at redshifts $z\lesssim2$. 
This LTB universe model has a void with size comparable to
the Hubble horizon scale. 
We first determine the decoupling epoch in this LTB universe model 
by an approximate analytical condition under a few simplified
assumptions on the physical quantities at that epoch. 
Then we calculate the cosmic microwave background (CMB) anisotropy
observed in the rest frame of clusters of galaxies 
which are assumed to be at rest in the spatial comoving coordinates 
of the LTB universe model. 
We find that the obtained temperature anisotropies are 
dominated by dipole, although there may exist higher multi-poles
in general. We may interpret this dipole anisotropy  
as the drift velocity of a cluster of galaxies
relative to the CMB rest frame. Hence it gives rise to
the kSZ effect. We calculate this effect and 
compare it with observational data. We find that
if we assume the conventional adiabatic perturbation scenario
at the time of decoupling, the drift velocity of clusters
of galaxies becomes unacceptably large. 
Conversely, this observational constraint may be relaxed
by introducing a non-adiabatic (i.e., primordially isocurvature) 
component of inhomogeneities at the time of decoupling. 
However, our result indicates that the necessary isocurvature
perturbation amplitude is very large.
}

\begin{document} 


\section{Introduction}
\label{sec:intro}
Copernican principle, i.e., the assumption that we are not at a special 
place in the universe,  
plays a very crucial role in the modern physical cosmology. 
If we accept this principle, 
the observed highly isotropic distribution of the cosmic microwave 
background (CMB), except for its dipolar anisotropy, implies 
the homogeneity and isotropy of our universe.
The standard cosmological model based on 
this homogeneity and isotropy and on general relativity,
namely the $\Lambda$CDM model, naturally explains
various important observational facts, supporting strongly 
the validity of the Copernican principle, though 
the homogeneity of our universe has not yet been directly tested. 

Within the framework of homogeneous and isotropic cosmological models, 
the distance-redshift relation of type Ia supernovae and the 
CMB data by WMAP lead us to a conclusion
that there is so-called dark energy as a major component of the universe. 
The present observational data is consistent with the dark energy
being a positive cosmological constant.
However, the energy density of the dark energy is about $10^{-120}$ 
times smaller than the Planck energy density. This number is 
regarded as one of the most unnatural numbers of the universe.
Hence, recently, many authors have been discussing the possibility 
to explain observations without dark energy by 
using anti-Copernican universe models in which we are assumed to be
at a very special position in the universe.
Usually, in anti-Copernican models, we are assumed to be located at
at the center of a spherically symmetric inhomogeneous 
universe~\cite{Zehavi:1998gz,
Tomita:1999qn,Tomita:2000jj,Tomita:2001gh,Celerier:1999hp}. 

One of the important attempts in this direction is 
to explain the type Ia supernovae observation using 
spherically symmetric inhomogeneities without dark energy.
There exits an exact solution of the Einstein equations known as 
the Lema\^itre-Tolman-Bondi (LTB) solution which describes 
a system of spherically symmetric dust. 
Actually, it is possible to construct an LTB 
universe model whose distance-redshift relation observed at 
the symmetry center 
agrees with that of the concordance $\Lambda$CDM 
model~\cite{Celerier:1999hp,Mustapha:1998jb,Iguchi:2001sq,Yoo:2008su,Celerier:2009sv,Kolb:2009hn}.  
Such an LTB universe model has a very large void structure comparable to 
the cosmological horizon scale. This result is very important 
because it implies that, in order to know the precise information about 
the dark energy, we have to know how to distinguish the effect of 
the dark energy from that of inhomogeneity. For this purpose, 
in this paper, we also 
study the LTB cosmology with an observer like us at its center. 

CMB observations in LTB universe models are also often 
discussed~\cite{Alnes:2006pf,Alnes:2005rw,Alexander:2007xx,Bolejko:2008cm,Clarkson:1999yj,Caldwell:2007yu,GarciaBellido:2008gd,GarciaBellido:2008nz,GarciaBellido:2008yq,Goodman:1995dt,Godlowski:2004gh,Zibin:2008vj,Zibin:2008vk,Clifton:2009kx,Regis:2010iq,Kodama:2010gr,Garfinkle:2009uf,Clarkson:2010ej,Moss:2010jx,Biswas:2010xm}. 
In many of these works, it is assumed that the universe 
is homogeneous in the spacelike asymptotic region 
from which the CMB photons come, hence
the CMB photon distribution is assumed to
be the same as that in the homogeneous and isotropic 
universe models at the last scattering surface (LSS). 

One of the most stringent constraints on the LTB universe model  
with a large void come from the observation of the 
kinematic Sunyaev-Zel'dovich (kSZ) effect. 
In the case of a homogeneous and isotropic universe, 
if a cluster of galaxies has a finite drift velocity
relative to the CMB comoving frame, 
it causes a dipole anisotropy in the CMB temperature 
in the cluster rest frame.
Then Compton scattering of the CMB photons
inside a cluster of galaxies causes a distortion
in the CMB spectrum in the direction of the cluster
of galaxies. Therefore observation of distortions in the
CMB spectrum constrains the drift velocity of
clusters of galaxies.

In the case of LTB universe models, 
even if clusters of galaxies are comoving 
with the non-relativistic matter component  
in the universe, or in other words, even if they
are at rest in the comoving spatial coordinates, 
an anisotropy of the CMB temperature will be observed at 
an off-center cluster due to the inhomogeneity of the universe. 
In contrast to a homogeneous and isotropic 
universe, this effect may not necessarily be of kinematic origin, 
since higher multi-pole components of the CMB anisotropy 
may exist in the cluster rest frame. 
This is because an LTB universe model 
is neither homogeneous nor isotropic at off-center clusters of galaxies. 
In this sense, the terminology `the kSZ effect' is not so appropriate 
for this effect in the LTB universe model.  
However, as will be shown later, in our LTB universe 
model, the anisotropy observed in the rest frame of
a cluster of galaxies is found to be dominated by 
a dipole component. Since a dipole anisotropy may be
regarded as the effect of the drift velocity of a cluster of galaxies,
we may call this effect the kSZ effect.

In Ref.~\cite{GarciaBellido:2008gd}, 
Garcia-Bellido and Haugboelle reported that 
current observations of only nine clusters with large error bars 
would already rule out an LTB universe model with a void of a radius 
greater than $\sim 1.5$Gpc. 
Their LTB universe model also has a large spherical void 
structure, but it is homogeneous and isotropic in the spatially 
asymptotic region. 
Thus, the CMB at and near the last scattering surface 
is estimated by the same procedure as in the case of a
 homogeneous and isotropic universe model. 
Differently from standard cosmology,
however, in the LTB universe models, it might be possible to 
introduce radial inhomogeneities in the CMB temperature.

In this paper, we analyze the kSZ effect in the 
LTB universe model whose distance-redshift relation observed 
at the center agrees with the concordance $\Lambda$CDM model,  
and then we show that the observational constraints from the kSZ effect 
have ruled out LTB universe models based on the adiabatic perturbation
scenario, but not so if we allow the existence of
non-adiabatic inhomogeneities at the decoupling epoch. 

This paper is organized as follows. 
In \S\ref{sec:LSS}, we introduce the LTB solution 
and show how to fix the decoupling epoch in an LTB universe model. 
Under some reasonable assumptions on the physical quantities at 
the decoupling epoch, 
we calculate the dipole anisotropy of the CMB temperature observed 
at each cluster of galaxies in the adiabatic scenario in which
there existed only adiabatic curvature perturbations 
in the early universe in \S\ref{sec:simplest}. We find that
the dipole anisotropy is too large to be consistent with observational data.
Then in \S\ref{sec:fidcl}, we assume the existence of
isocurvature inhomogeneities at the decoupling epoch, and show 
that it may save LTB universe models.
Finally, \S\ref{sec:sudi} is devoted to summary and discussion. 

\section{Decoupling Epoch in the LTB Universe}
\label{sec:LSS}
\subsection{Background universe model}
\label{sec:background}
In Ref.~\cite{Yoo:2008su}, Yoo, Kai and Nakao 
numerically constructed 
an LTB universe model whose distance-redshift relation 
agrees with that of the concordance $\Lambda$CDM model 
in the whole redshift domain and which 
is homogeneous at the early stage of the universe. 
Recently we modified this LTB universe model 
at redshifts $z\gtrsim2$ so that 
the peak positions in the CMB temperature anisotropy 
are consistent with the observation~\cite{Yoo:2010qy}
 (hereafter referred to as Paper I).
In this paper, we use this LTB model as the background 
universe. 

The metric of the LTB solution is given by
\begin{equation}
ds^2=-c^2dt^2+\frac{\left(\partial_r R(t,r)\right)^2}{1-k(r)r^2}dr^2
+R^2(t,r)(d\vartheta^2+\sin^2\vartheta d\varphi^2), \label{eq:metric}
\end{equation}
where $k(r)$ is an arbitrary function of the radial coordinate $r$. 
The matter is dust whose stress-energy tensor is given by
\begin{equation}
T^{\mu\nu}=\rho u^\mu u^\nu,
\end{equation}
where $\rho=\rho(t,r)$ is the 
mass density, 
and $u^\mu$ is the four-velocity of 
the fluid element. 
The coordinate system in Eq.~(\ref{eq:metric}) is chosen
in such a way that $u^\mu=(1,0,0,0)$.

The area radius $R(t,r)$ 
satisfies one of the Einstein equations, 
\begin{equation}
\left(\frac{\partial R}{\partial t}\right)^2=\frac{2GM(r)}{R}-c^2k(r)r^2,
\label{eq:Einstein-eq}
\end{equation}
where $M(r)$ is an arbitrary function related to 
the 
mass 
density $\rho$ by
\begin{equation}
\rho(t,r)=\frac{1}{4\pi R^2(t,r)}\frac{dM(r)}{dr}.
\end{equation}
Following \cite{Tanimoto:2007dq}, we write the solution 
of Eq.~(\ref{eq:Einstein-eq}) in the form,
\begin{eqnarray}
R(t,r)&=&(6GM(r))^{1/3}(t-t_{\rm B}(r))^{2/3} S(x), 
\label{eq:YS}\\
x&=&c^2k(r)r^2\left(\frac{t-t_{\rm B}(r)}{6GM(r)}\right)^{2/3},
 \label{eq:defx}
\end{eqnarray}
where $t_{\rm B}(r)$ is an arbitrary function 
which determines the big bang time, 
and $S(x)$ is a function defined implicitly as
\begin{equation}
S(x)=
\left\{\begin{array}{lll}
\displaystyle
\frac{\cosh\sqrt{-\eta}-1}{6^{1/3}(\sinh\sqrt{-\eta}
-\sqrt{-\eta})^{2/3}}
\,;\qquad
&\displaystyle
x=\frac{-(\sinh\sqrt{-\eta}-\sqrt{-\eta})^{2/3}}{6^{2/3}}
\quad&\mbox{for}~~x<0\,,
\\
\displaystyle
\frac{1-\cos\sqrt{\eta}}{6^{1/3}(\sqrt{\eta}
-\sin\sqrt{\eta})^{2/3}}
\,;&\displaystyle
x=\frac{(\sqrt{\eta}-\sin\sqrt{\eta})^{2/3}}{6^{2/3}}
\quad&\mbox{for}~~x>0\,,
\end{array}\right.
\label{eq:defS}
\end{equation}
and $S(0)=({3}/{4})^{1/3}$. 
The function $S(x)$ is analytic for $x<(\pi/3)^{2/3}$. 
Some characteristics of the function $S(x)$ 
are given in \cite{Tanimoto:2007dq} and \cite{Yoo:2008su}. 

As shown in the above, the LTB solution has three arbitrary functions, 
$k(r)$, $M(r)$ and $t_{\rm B}(r)$. 
One of them is a gauge degree of freedom 
for rescaling of the radial coordinate $r$.
We fix this by setting
\begin{equation}
M(r)=\frac{4}{3}\pi\rho_0r^3, 
\end{equation}
where $\rho_0$ is the energy density at the 
center at present $\rho_0=\rho(t_0,0)$.
As in the case of the homogeneous and isotropic 
universe, the present Hubble parameter $H_0$ is related to $\rho_0$ as 
\begin{equation}
H_0^2+k(0)c^2=\frac{8}{3}\pi G\rho_0. 
\end{equation}
As in \cite{Yoo:2008su}, we assume the simultaneous big bang, i.e., 
\begin{equation}
t_{\rm B}(r)=0. \label{eq:tB}
\end{equation}
For notational simplicity, we introduce 
dimensionless quantities,
\begin{eqnarray*}
\tilde r:=\frac{H_0r}{c}\,,\quad \tilde k(\tilde r):=\frac{k(r)c^2}{H_0^2}\,. 
\end{eqnarray*}

The LTB universe model proposed in Paper I is 
specified by the following curvature function:
\begin{equation}
\widetilde k(\widetilde r)=\widetilde k_{\rm fit}
(\widetilde r)\times f(\widetilde r;A), 
\end{equation}
where  
\begin{eqnarray}
\widetilde k_{\rm fit}(\widetilde r)&=&
\frac{0.545745}{0.211472+ \sqrt{0.026176+ \tilde r}} 
- \frac{2.22881}{\left(0.807782+ \sqrt{0.026176+ \tilde r}\right)^2},\\
f(\widetilde r;A)&=&
\left\{
\begin{array}{lll}
1~~&{\rm for}~~\widetilde r<2\\
1+\frac{16A\left(\widetilde r-2\right)^3\left(323-123\widetilde r
+12\widetilde r^2\right)}{3125}~~&{\rm for}~~2\leq \widetilde r<9/2\\
1+A~~&{\rm for}~~9/2\leq \widetilde r 
\end{array}
\right. 
\end{eqnarray}
with $A=-1.069$. 
The distance-redshift relation of this LTB solution agrees with that of 
the concordance $\Lambda$CDM model at redshifts $z\lesssim2$.

\subsection{Decoupling epoch in the LTB universe}
\label{sec:LSSinLTB}
The decoupling between photons and baryons occurs 
in an inhomogeneous universe just as in the case of
a homogeneous universe. That is, it occurs when the mean free path 
of photons becomes effectively infinite due to almost
complete recombination of electrons to protons.

Since there is no radiation component in our LTB model,
we cannot treat the decoupling in a rigorous manner.
However, as in the concordance model, we expect
the energy density of the radiation to be only a small 
fraction of the total density at decoupling, hence its effect
on the spacetime geometry is small, if not negligible. 
In fact, the radiation energy density estimated in our LTB model
turns out to be about 20\% of the total energy density.
This means that treating the radiation as a test field in
our model is consistent to a first approximation.

\begin{figure}[htbp]
\begin{center}
\includegraphics[scale=0.7]{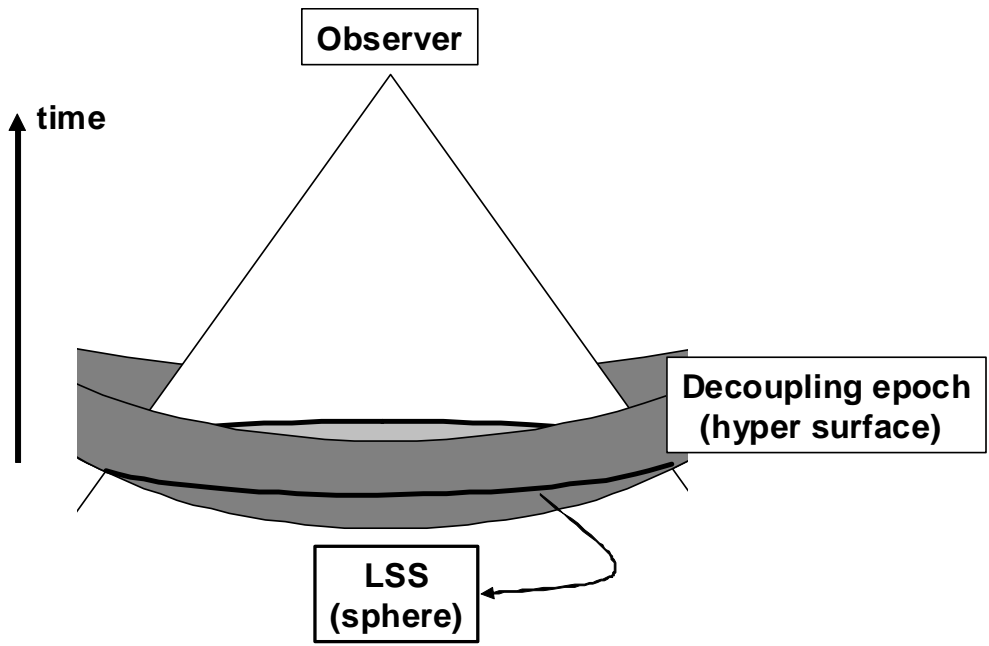}
\caption{Schematic figure for the hypersurface of the decoupling epoch 
in an LTB universe. 
}
\label{fig:lightconeandlss}
\end{center}
\end{figure}

Another approximation we adopt is the instantaneous decoupling.
Namely, we assume decoupling to occur on a single spacelike
hypersurface. 
Since our LTB universe model is inhomogeneous but spherically symmetric,
it is natural to assume that the decoupling hypersurface is also 
inhomogeneous but spherically symmetric.
Thus it is specified by the form,
\begin{equation}
t=t_{\rm D}(r)\,.
\end{equation}
The cross section of this hypersurface with
the past directed null cone from the observer at the center
constitutes the LSS of CMB photons
 (see Fig.~\ref{fig:lightconeandlss}).
The LSS is a spacelike 2-dimensional sphere 
by the assumed symmetry.  
We use the subscript $*$ to express physical quantities 
on the LSS. In our approximation, ignoring secondary effects, 
the CMB anisotropy is essentially determined by the distribution
of photons on the LSS.

The geodesic equations to determine the past light cone from
the observer at the center are written in the form,
\begin{eqnarray}
(1+z)\frac{dt}{dz}
&=&-\frac{\partial _r R}{\partial_t\partial_rR},
\label{eq:nullgeo1}\\
(1+z)\frac{dr}{dz}
&=&\frac{c\sqrt{1-k(r)r^2}}{\partial_t\partial_rR}, 
\label{eq:nullgeo2}
\end{eqnarray}
where the past directed radial null geodesics have been 
parametrized by the cosmological redshift $z$.
We denote the solution of the above equations by
\begin{equation}
t=t_{\rm lc}(z)~,~r=r_{\rm lc}(z). 
\end{equation}

Now to discuss the decoupling condition, 
for simplicity, we consider a universe consists of
cold dark matter, protons and electrons, and neutral hydrogen atoms.
In particular, we neglect helium. Since the contributions of
helium and the other components are not large, this simplification 
should not lead to a serious error in our analysis.
We also assume that, until the decoupling time,
photons, electrons and protons are in thermal equilibrium. 
The energy density of the electrons and photons is negligible, and hence 
the constituents of our LTB model are cold dark matter 
and baryons, the latter of which consist of protons and hydrogen atoms.
Thus the baryon number density 
$n_{\rm b}$ is equal to the total number density of 
protons and hydrogen atoms, 
and the electron number density $n_{\rm e}$ is equal to
the proton number density.

In homogeneous and isotropic cosmology, 
the decoupling time is well determined by Gamow's criterion,
\begin{equation}
H= \Gamma,
\label{eq:Gamow1}
\end{equation}
where $H$ is the Hubble parameter and $\Gamma$ is the rate
of collisions of a photon with electrons.
Using the Thomson scattering cross section $\sigma_{\rm T}$ and the 
electron number density $n_{\rm e}$, $\Gamma$ is written as
\begin{equation}
\Gamma=cn_{\rm e}\sigma_{\rm T}\,.
\end{equation}
In our LTB model, we adopt this criterion,
with the identification of the``Hubble parameter $H$'' 
with 
\begin{equation}
H^2=\frac{8\pi G}{3}\rho\,.
\end{equation}
Since, by virtue of Eq.~(\ref{eq:tB}), our LTB universe model 
is almost identical to the Einstein-de Sitter universe near the 
decoupling time. Hence the above definition of $H$
is accurate enough for our purpose. 

In order to estimate the electron number density, 
for simplicity, we use Saha's equation 
assuming thermal equilibrium. 
We note that our assumption of thermal equilibrium 
at the decoupling epoch is not appropriate in reality, 
because the decoupling process is controlled by a 
non-thermal physical process~\cite{Weinberg:2008zzc}. 
However, as mentioned in \cite{Clarkson:2010ej}, 
the error in the temperature estimate turns out to be 2 - 3\%. 
Hence our treatment of the decoupling is good enough for
our qualitative discussions about the kSZ effect. 

Let us consider the ionization rate $X_{\rm e}:=n_{\rm e}/n_{\rm b}$. 
In thermal equilibrium, the ionization rate $X_{\rm e}$ satisfies 
Saha's equation,
\begin{equation}
\frac{1-X_{\rm e}}{X_{\rm e}^2}=\frac{4\sqrt{2}\zeta(3)}{\sqrt{\pi}}\eta
\left(\frac{k_{\rm B}T}{m_{\rm e}c^2}\right)^{3/2}
\exp\left(\frac{13.59 {\rm eV}}{k_{\rm B}T}\right), 
\end{equation}
where $\zeta(x)$ is the zeta function, and 
$T$, $\eta$, $k_{\rm B}$ and $m_{\rm e}$ are the temperature, 
the baryon-to-photon ratio, 
the Boltzmann constant and the electron mass, respectively. 
Since the ionization rate at decoupling drops down to
$X_e\sim 10^{-5}$, we may approximate the above equation by 
\begin{equation}
X_{\rm e}^2\simeq\frac{\sqrt{\pi}}{4\sqrt{2}\zeta(3)}\frac{1}{\eta}
\left(\frac{k_{\rm B}T}{m_{\rm e}c^2}\right)^{-3/2}
\exp\left(-\frac{13.59{\rm eV}}{k_{\rm B}T}\right). 
\end{equation}
Using this equation, Gamow's criterion (\ref{eq:Gamow1}) 
is rewritten in the form,
\begin{equation}
\eta=\frac{32\sqrt{2\pi}\zeta(3)}{3} 
\frac{G\rho}{(cn_{\gamma 0}\sigma_{\rm T})^2}
\left(\frac{k_{\rm B}T_0}{m_{\rm e}c^2}\right)^{3/2}
\left(\frac{T_0}{T}\right)^{9/2}
\exp\left(\frac{13.59{\rm eV}}{k_{\rm B}T}\right), 
\label{eq:Gamow2}
\end{equation}
where $n_{\gamma0}$ and $T_0\simeq 2.725$K are the present 
photon number density and observed CMB temperature, respectively. 

Since we assume thermal equilibrium of electrons, 
protons and photons until the decoupling time $t=t_{\rm D}(r)$, 
the physical state of the decoupling hypersurface,
which is spherically symmetric, is determined by the 
distributions of the temperature $T=\Tlss(r)$, 
the baryon-to-photon ratio  $\eta=\etalss(r)$,
and the matter energy density $\rho=\rho(t_{\rm D}(r),r)$.
For convenience, in place of $\rho(t_{\rm D}(r),r)$,
we introduce the following quantity:
\begin{equation}
\alphalss(r):=\frac{\rho(t_{\rm D}(r),r)}{\rho_0}
\left(\frac{T_0}{\Tlss(r)}\right)^3. \label{eq:alpha}
\end{equation}
The quantity $\alphalss(r)$ is proportional 
to the ratio of the matter density 
and the photon number density. Then, from Eq.~(\ref{eq:Gamow2}), we obtain 
\begin{equation}
\etalss(r)=\frac{32\sqrt{2\pi}\zeta(3)}{3} 
\frac{G\rho_0}{(cn_{\gamma 0}\sigma_{\rm T})^2}
\left(\frac{k_{\rm B}T_0}{m_{\rm e}c^2}\right)^{3/2}
\alphalss(r)
\left(\frac{T_0}{\Tlss(r)}\right)^{3/2}
\exp\left(\frac{13.59{\rm eV}}{k_{\rm B}\Tlss(r)}\right).
\label{eq:gamow}
\end{equation}
We note that once $\Tlss(r)$ and $\alphalss(r)$ are given, 
the hypersurface $t=t_{\rm D}(r)$ can be obtained 
from Eq.~(\ref{eq:alpha}). 

On the LSS, we must have
\begin{equation}
\frac{\Tlss(r_{\rm lc}(z_*))}{T_0}
=\frac{T_*}{T_0}=1+z_*. 
\label{eq:zlss}
\end{equation}
Then, if we regard $\Tlss(r)$, $\etalss(r)$ and $\alphalss(r)$ 
as mutually independent functions, 
we have one 
functional condition (\ref{eq:gamow}) and 
one boundary condition at $z=z_*$ to 
constrain these three functions. 
But these are not enough to determine the decoupling hypersurface, 
$t=t_{\rm D}(r)$, through Eq.~(\ref{eq:alpha}). 
We would need two more functional conditions.
However, to know the location of the LSS for the 
observer at the center, 
we only need a single condition on the LSS 
in addition to Eqs.~(\ref{eq:gamow}) and (\ref{eq:zlss}).
For simplicity, we impose 
\begin{equation}
\eta_*=6.2\times10^{-10}\, \label{eq:eta-star}
\end{equation}
as in Paper I~\cite{Yoo:2010qy}.

We numerically solve the radial null geodesic 
equations (\ref{eq:nullgeo1}) and (\ref{eq:nullgeo2}). 
At each redshift $z$, we define $T$ and $\alpha$ by  
\begin{eqnarray}
T&=&(1+z)T_0,\\
\alpha&=&\frac{\rho(t_{\rm lc}(z),r_{\rm lc}(z))}{\rho_0(1+z)^3}. 
\end{eqnarray}
During the numerical integration,
we check at each redshift $z$ whether Eq.~(\ref{eq:gamow}) 
is satisfied by $\Tlss=T$, $\alphalss=\alpha$ 
and $\etalss=\eta_*$. 
If Eq.~(\ref{eq:gamow}) is satisfied, we 
stop integrating Eqs.~(\ref{eq:nullgeo1}) and (\ref{eq:nullgeo2}), 
and identify $z$, $T$ and $\alpha$ 
at this moment with $z_*$, $T_*$ and $\alpha_*$, 
respectively. 
For our LTB universe model, we have obtained
\begin{eqnarray}
\frac{T_*}{T_0}&=&1+z_*\simeq1130\,, 
\label{eq:T-result}\\
\alpha_*&\simeq&5.184\,, \label{eq:alpha-result}
\end{eqnarray}

\section{Kinematic Sunyaev-Zel'dovich effect}

\begin{figure}[htbp]
\begin{center}
\includegraphics[scale=0.7]{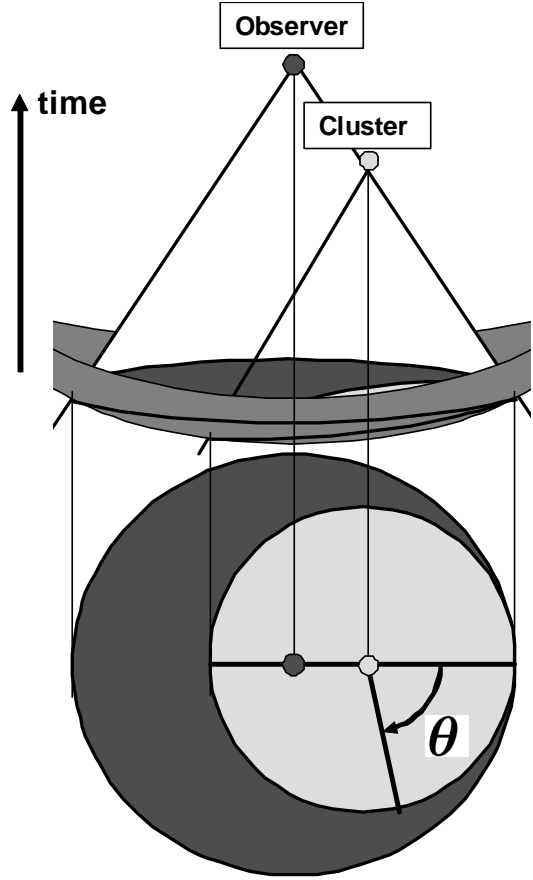}
\caption{A schematic figure of the spacetime and the LSS. 
}
\label{fig:lightcone}
\end{center}
\end{figure}

We consider the kSZ effect caused by the hot gases 
bound within clusters of galaxies which are assumed to be
at rest in the spatial coordinates of the LTB universe model. 
Let us consider a cluster of galaxies at redshift $z=\zcl$. 
In order to know how the CMB is observed at the position
of this cluster of galaxies, 
we numerically integrate pastward the null geodesic equations 
from this cluster to every direction (see Fig.~\ref{fig:lightcone}). 
Without loss of generality, 
we may focus on null geodesics in the plane $\varphi=0$.
Then the non-radial null geodesic equations are given by 
\begin{eqnarray}
t''&=&-\frac{(\partial_r R)\partial_t\partial_r R}{1-kr^2}~{r'}^2
-R(\partial_t R)~{\vartheta'}^2, 
\label{eq:nonradnull1}
\\
r''&=&-2\left[\partial_t(\ln\partial_r R)\right]~t'r'
+\frac{1}{2}\left[\partial_r\ln(1-kr^2)
-2\partial_r(\ln\partial_r R)\right]~{r'}^2
+\frac{(1-kr^2)R}{\partial_r R}~{\vartheta'}^2, 
\label{eq:nonradnull2}
\\
\vartheta''&
=&2(\partial_t\ln R)~t'\vartheta'-2(\partial_r\ln R)~r'\vartheta', 
\label{eq:nonradnull3}
\end{eqnarray}
where the prime denotes differentiation with an affine parameter. 
From the null condition, we have 
\begin{equation}
{r'}^2=\frac{1-kr^2}{\left(\partial_rR\right)^2}
\left({t'}^2-R^2~{\vartheta'}^2\right).
\label{eq:nonradnullcon}
\end{equation}
The initial conditions 
$(t_{\rm i},r_{\rm i},\vartheta_{\rm i},t'_{\rm i},r'_{\rm i},\vartheta'_{\rm i})$ 
to integrate the null geodesic equations are given by 
\begin{eqnarray}
t_{\rm i}&=&t_{\rm c}:=t_{\rm lc}(z_{\rm c}), 
\label{eq:inicon1}
\\
r_{\rm i}&=&r_{\rm c}:=r_{\rm lc}(z_{\rm c}),
\label{eq:inicon2}
\\
\vartheta_{\rm i}&=&0,
\label{eq:inicon3}
\\
t'_{\rm i}&=&-(1+z_{\rm c}),
\label{eq:inicon4}
\\
r'_{\rm i}&=&\frac{\sqrt{1-k(r_{\rm c})r_{\rm c}^2}}{
\left.\partial_rR
\right|_{t=t_{\rm c},r=r_{\rm c}}}
(1+z_{\rm c})\cos\theta, 
\label{eq:inicon5}
\\
\vartheta'_{\rm i}&=&
\frac{1}{\left.R
\right|_{t=t_{\rm c},r=r_{\rm c}}}
(1+z_{\rm c})\sin\theta, 
\label{eq:inicon6}
\end{eqnarray}
where $\theta$ describes the direction of the null geodesic 
in the cluster rest frame (see Fig.~\ref{fig:lightcone}).  
Note that Eq.~(\ref{eq:inicon4}) is a condition to 
fix the degree of freedom for the affine transformation. 
It is easy to see that Eq.~(\ref{eq:nonradnullcon}) 
is satisfied by these initial conditions. 
The redshift at each point on the null geodesic 
is defined by 
\begin{equation}
1+z=-t'. 
\end{equation}
%

At the intersection between the past light-cone of the cluster 
of galaxies located at $z=z_{\rm c}$ and the 
decoupling hypersurface, 
we stop integrating Eqs.~(\ref{eq:nonradnull1})--(\ref{eq:nonradnull3}) 
and label the redshift at this intersection 
as $z_{\rm D}(\theta;\zcl)$. Here note that $z_{\rm D}(0;\zcl)=z_*$. 
Then, the CMB temperature $T_{\rm cl}(\theta;\zcl)$ 
observed at the cluster of galaxies located at $z=\zcl$ 
is obtained by 
\begin{equation}
T_{\rm cl}(\theta;\zcl)=\frac{1+\zcl}{1+z_{\rm D}(\theta;\zcl)}
T_{\rm D}\left(r_{\rm D}\left(\theta;\zcl\right)\right),
  \label{eq:Tc}
\end{equation}
where $r=r_{\rm D}(\theta;\zcl)$ is the radial coordinate of the intersection 
between the past light-cone of this cluster and the decoupling 
hypersurface. 
If we know the temperature $T_{\rm D}(r)$ on the decoupling 
hypersurface, we find $T_{\rm cl}(\theta;\zcl)$ by the above equation. 
Conversely, if we know the information about 
$T_{\rm cl}(\theta;\zcl)$ through the kSZ effect, 
we find $\Tlss(r)$ for the domain
 $r_{\rm D}(\pi;z_c)\leq r\leq r_{\rm D}(0;z_c)$. 

\subsection{ Case of homogeneous decoupling hypersurface}
\label{sec:simplest}
As mentioned in the second to the last paragraph in \S\ref{sec:LSSinLTB}, 
we do not have enough information to determine the physical quantities 
on the decoupling hypersurface, $\Tlss(r)$, $\etalss(r)$ and $\alphalss(r)$ 
and hence the embedding of the decoupling hypersurface $t=t_{\rm D}(r)$. 
To determine them, we need to impose two additional assumptions on
these quantities. 
Here, we assume $\Tlss(r)$ and $\etalss(r)$ are constant, 
i.e., $\Tlss(r)=T_*$ and $\etalss(r)=\eta_*$.  
Then $\alphalss(r)$ also becomes a constant determined by 
Eq.~(\ref{eq:gamow}). Consequently, from Eq.~(\ref{eq:alpha}), 
we can determine the decoupling hypersurface $t=t_{\rm D}(r)$. 

\begin{figure}[htbp]
\begin{center}
\includegraphics[scale=1.5]{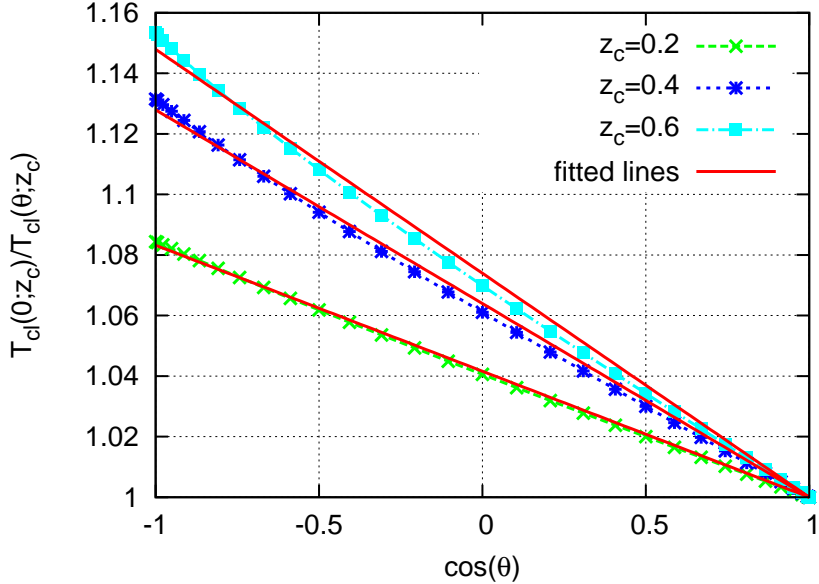}
\caption{
The temperature anisotropy seen at a cluster of galaxies, 
$T_{\rm cl}(\theta;z_{\rm c})$, and the dipole fit to it,
$T_{\rm dipole}(\theta;z_{\rm c})$, for several values of $z_c$.}
\label{fig:fits}
\end{center}
\end{figure}

Since an LTB universe model is neither homogeneous nor isotropic 
for an observer at an off-center cluster of galaxies, 
not only a dipole component but also 
higher multi-pole components in $T_{\rm cl}(\theta;\zcl)$ may exist. 
However, we find from our numerical results 
that the anisotropy in $T_{\rm cl}(\theta;\zcl)$ is dominated by 
dipole, and thus we can fit $T_{\rm cl}(\theta;\zcl)$ with 
the Doppler shifted temperature distribution,
\begin{equation}
T_{\rm dipole}(\theta;\zcl)
:=\frac{c-\vcl(\zcl)}{c-\vcl(\zcl) \cos\theta}\Tcl(0;\zcl) ,
\end{equation}
where $\vcl(\zcl)$ is regarded as 
the effective drift velocity of the cluster at $z=\zcl$ relative to 
the CMB rest frame, and is determined by using the 
least-square method (see Fig.~\ref{fig:fits}).

\begin{figure}[htbp]
\begin{center}
\includegraphics[scale=2.]{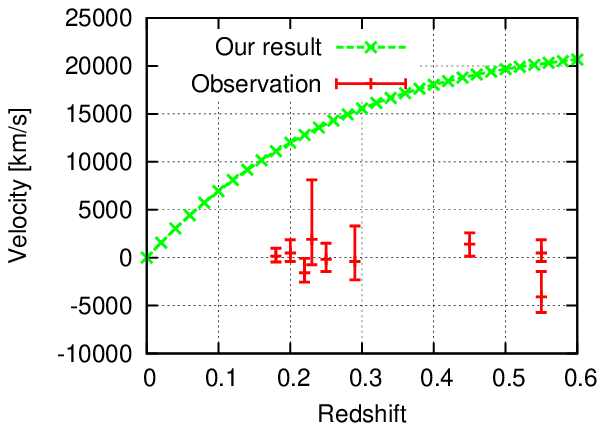}
\caption{Effective drift velocity at each cluster redshift
for a homogeneous temperature distribution 
on the decoupling hypersurface. 
}
\label{fig:vconsts}
\end{center}
\end{figure}

The result is shown in Fig.~\ref{fig:vconsts} compared with the 
observed values for nine clusters reported 
in \cite{Holzapfel:1997ui,Benson:2003va,Kitayama:2003mq}. 
The observational data are listed in Table~\ref{tab:clob}. 
\begin{table}[htbp]
\caption{Observed drift velocities for nine clusters. 
}
\label{tab:clob}
\begin{tabular}{|c|c|c|c|c|}
\hline
$i$(1 - 9): name&redshift($z_i$)&drift velocity[km/s]($v_i$)&$\sigma^+_i$[km/s]&$\sigma^-_i$[km/s]\\
\hline
\taboxl{1: A1689\cite{Holzapfel:1997ui}}
&\tabox{0.18}
&\tabox{+170}
&\tabox{+815}
&\tabox{$-$630}
\\
\hline
\taboxl{2: A2163\cite{Holzapfel:1997ui}}
&\tabox{0.20}
&\tabox{+490}
&\tabox{+1370}
&\tabox{$-$880}
\\
\hline
\taboxl{3: A2261\cite{Benson:2003va}}
&\tabox{0.22}
&\tabox{$-$1575}
&\tabox{+1500}
&\tabox{$-$975}
\\
\hline
\taboxl{4: A2396\cite{Benson:2003va}}
&\tabox{0.23}
&\tabox{+1900}
&\tabox{+6225}
&\tabox{$-$2650}
\\
\hline
\taboxl{5: A1835\cite{Benson:2003va}}
&\tabox{0.25}
&\tabox{$-$175}
&\tabox{+1675}
&\tabox{$-$1275}
\\
\hline
\taboxl{6: Zw 3146\cite{Benson:2003va}}
&\tabox{0.29}
&\tabox{$-$400}
&\tabox{+3700}
&\tabox{$-$1925}
\\
\hline
\taboxl{7: RX J1347-1145\cite{Kitayama:2003mq}}
&\tabox{0.45}
&\tabox{+1420}
&\tabox{+1170}
&\tabox{$-$1270}
\\
\hline
\taboxl{8: Cl 0016 + 16\cite{Benson:2003va}}
&\tabox{0.55}
&\tabox{$-$4100}
&\tabox{+2650}
&\tabox{$-$1625}
\\
\hline
\taboxl{9: MS 0451\cite{Benson:2003va}}
&\tabox{0.55}
&\tabox{+490}
&\tabox{+1370}
&\tabox{$-$880}
\\
\hline
\end{tabular}
\end{table}
%
Obviously, the effective drift velocity in our model is 
far larger than the observed values. 
This fact means that our LTB universe model 
whose $\Tlss(r)$, $\etalss(r)$ and $\alphalss(r)$ are constant 
cannot explain the observational data. 

Here we have assumed that the ratios of two of any physical quantities 
(the energy densities of baryon, dark matter, photons,
temperature of photon) are constant on the decoupling hypersurface. 
Since, by virtue of the assumption (\ref{eq:tB}), our background
LTB universe model is very close to a homogeneous Einstein-de Sitter 
universe at the decoupling epoch, these assumptions imply that
inhomogeneities are small and adiabatic at that time.
Therefore the above result strongly indicates that anti-Copernican 
LTB cosmological models are virtually excluded by 
the observational data within the context of the conventional 
adiabatic perturbation scenario for the structure formation.

In \cite{GarciaBellido:2008gd,Garfinkle:2009uf},  
homogeneity at the last scattering surface was also assumed 
and their results are qualitatively consistent with 
the result obtained in this section. 


\subsection{Case of inhomogeneous  decoupling hypersurface}
\label{sec:fidcl}
We now consider the possibility of an inhomogeneous LSS
and look for an $r$-dependent decoupling hypersurface $t=t_{\rm D}(r)$ 
and the physical quantities $\Tlss(r)$, $\etalss(r)$ and $\alphalss(r)$ 
so that the observational results of the kSZ effect 
can be explained by our LTB universe model. 

\begin{figure}[htbp]
\begin{center}
\includegraphics[scale=0.7]{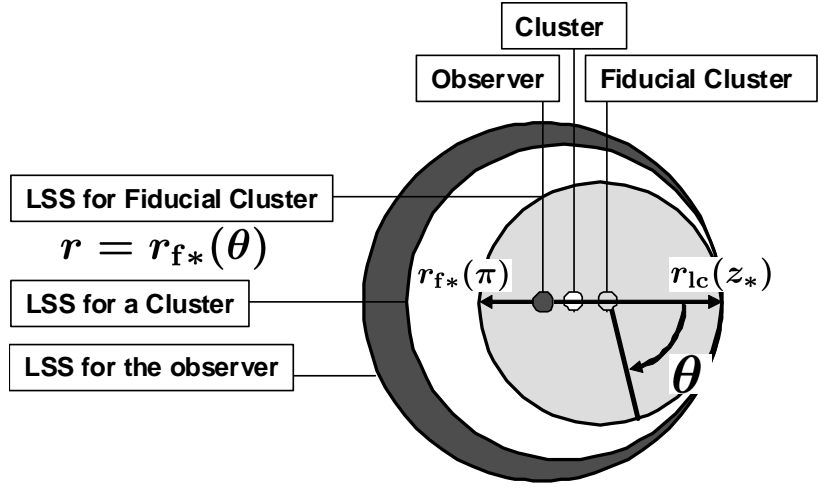}
\caption{Schematic figure of LSSs on the decoupling hypersurface
for the fiducial cluster, a cluster at a redshift $\zcl<z_{\rm f}$ 
and the observer at $z=0$ (us). 
}
\label{fig:lsss}
\end{center}
\end{figure}

First, for simplicity, we assume $\etalss(r)=\eta_*$
(see \cite{Regis:2010iq} about inhomogeneity of $\etalss$). 
Then, to fix the decoupling epoch, 
we consider a comoving fiducial cluster of galaxies 
at redshift $z_{\rm c}=z_{\rm f}$. 
We assume that the anisotropy of the CMB temperature observed 
in the rest frame of the fiducial cluster is purely dipolar, i.e., 
\begin{equation}
T_{\rm cl}(\theta;z_{\rm f})
=\frac{c-v_{\rm f}}{c-v_{\rm f}\cos\theta}T_{\rm cl}(0;z_{\rm f})
=\frac{c-v_{\rm f}}{c-v_{\rm f}\cos\theta}(1+z_{\rm f})T_0\,, 
\end{equation}
where $v_{\rm f}$ is the effective drift velocity of the 
fiducial cluster of galaxies relative to the CMB rest frame. 
Then, the decoupling hypersurface, $t=t_{\rm D}(r)$ is
obtained from the above CMB temperature $T_{\rm cl}(\theta;z_{\rm f})$ 
in the following manner:
We numerically integrate null 
geodesic equations (\ref{eq:nonradnull1})--(\ref{eq:nonradnull3}) 
pastward from this fiducial cluster of galaxies to every direction. 
At each redshift $z$ on the null geodesic to the direction 
specified by $\theta$, we define $T$ and $\alpha$ by 
\begin{eqnarray}
T&=&\frac{1+z}{1+z_{\rm f}}T_{\rm cl}(\theta;z_{\rm f}), 
\\
\alpha&=&\frac{\rho(t_{\rm g}(z;\theta),
r_{\rm g}(z;\theta))}{\rho_0}
\left(\frac{T_0}{T}\right)^3, \label{eq:alphageo}
\end{eqnarray}
where $t=t_{\rm g}(z;\theta)$ and $r=r_{\rm g}(z;\theta)$ 
represent the trajectory of this null geodesic (see Fig.~\ref{fig:lsss}). 

During the numerical integration,
we check at each redshift $z$ whether Eq.~(\ref{eq:gamow}) 
is satisfied by 
$\Tlss=T$, $\alphalss=\alpha$ 
and $\etalss=\eta_*$. 
If Eq.~(\ref{eq:gamow}) is satisfied, we 
stop integrating Eqs.~(\ref{eq:nonradnull1})--(\ref{eq:nonradnull3}). 
At this moment, we have $z=z_{\rm D}(\theta;z_{\rm f})$ and 
$r_{\rm g}(z;\theta)=r_{\rm D}(\theta;z_{\rm f})$.
This determines $T$ and $\alpha$ as functions of
$r$: $T=T_{\rm D}(r_{\rm D}(\theta;z_{\rm f}))$ and 
$\alpha=\alpha_{\rm D}(r_{\rm D}(\theta;z_{\rm f}))$, respectively. 
The schematic figure for the trajectory $r=r_{\rm D}(\theta;z_{\rm f})$ 
is shown in Fig.~\ref{fig:lsss}. 
We find from this figure that 
\begin{equation}
r_{\rm D}(\pi;z_{\rm f})
\leq r_{\rm D}(\theta;z_{\rm f})\leq r_{\rm D}(0;z_{\rm f})=\rlc(z_*)
~~{\rm for}~~0\leq\theta\leq\pi. 
\end{equation}
The above inequality means that 
the functional forms of $\Tlss(r)$ and $\alphalss(r)$ 
are determined only for the range 
 $r_{\rm D}(\pi;z_{\rm f})\leq r\leq \rlc(z_*)$. 
However, this information is enough for the calculation of the 
temperature anisotropies from other clusters at redshift $\zcl<z_{\rm f}$,
because the past light-cone emanated from this cluster 
intersects with the decoupling hypersurface  
inside the region $r_{\rm D}(\pi;z_{\rm f})\leq r\leq \rlc(z_*)$ 
as shown in Fig.~\ref{fig:lsss}. 

\begin{figure}[htbp]
\begin{center}
\includegraphics[scale=1.4]{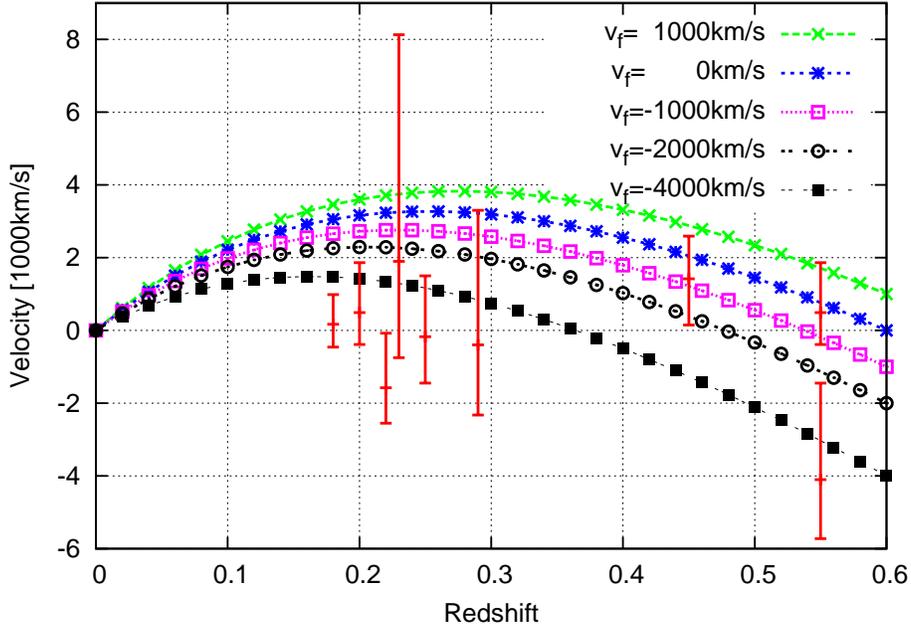}
\caption{The effective velocities as functions of the 
cluster redshift $\zcl$ are shown for various values of
$v_{\rm f}$. The fiducial cluster is at $z_{\rm f}=0.6$.
}
\label{fig:v}
\end{center}
\end{figure}

Applying the same procedure as in the previous section, 
we obtain the CMB temperature 
$T_{\rm cl}(\theta;z_{\rm c})$ observed at a cluster of galaxies located  
at $z=z_{\rm c}<z_{\rm f}$. Results are shown in Fig.~\ref{fig:v}, 
where we set $z_{\rm f}=0.6$.  
The effective velocities $v_{\rm c}(\zcl)$ 
for various
values of $v_{\rm f}$ are shown as functions of the 
cluster redshift $\zcl$. 
The effective velocity of a cluster
is evaluated from the temperature anisotropy 
seen from the cluster using the least-square method. 
It is clear that the result has much better 
consistency with the observational data than the model with an 
adiabatic situation in \S\ref{sec:simplest}. 

In order to compare our result with that obtained in 
\cite{GarciaBellido:2008gd}, we have calculated the same 
log-likelihood as theirs,
\begin{equation}
-2\ln \mathcal L=\sum_i\frac{
\left(v_i-\vcl(z_i)+v_{\rm sys}\right)^2}
{\sigma_i^{\pm2}+\sigma_{\rm pv}^2} \,.
\end{equation}
\begin{figure}[htbp]
\begin{center}
\includegraphics[scale=1.5]{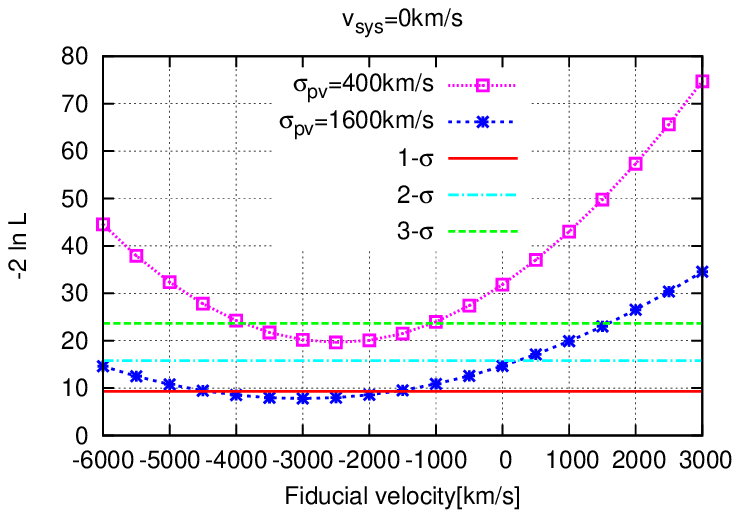}
\includegraphics[scale=1.5]{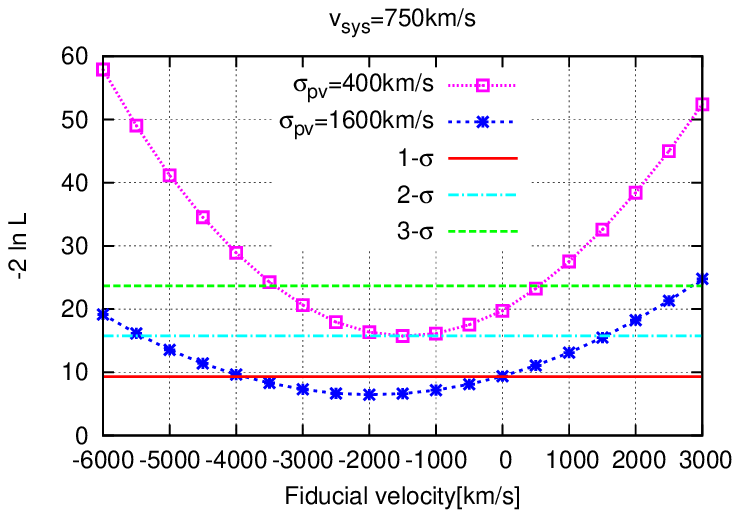}
\caption{Log-likelihoods for 
$(\sigma_{\rm pv}, v_{\rm sys})=(400,0)$, $(400,750)$, $(1600,0)$ and 
$(1600,750)$. 
}
\label{fig:loglikeli}
\end{center}
\end{figure}
As shown in the cases $\sigma_{\rm pv}=1600$km/s, 
even though our model has an extremely large void, 
it turns out to be consistent with the observational data
in the 1-$\sigma$ level. 

\begin{figure}[htbp]
\begin{center}
\includegraphics[scale=1.5]{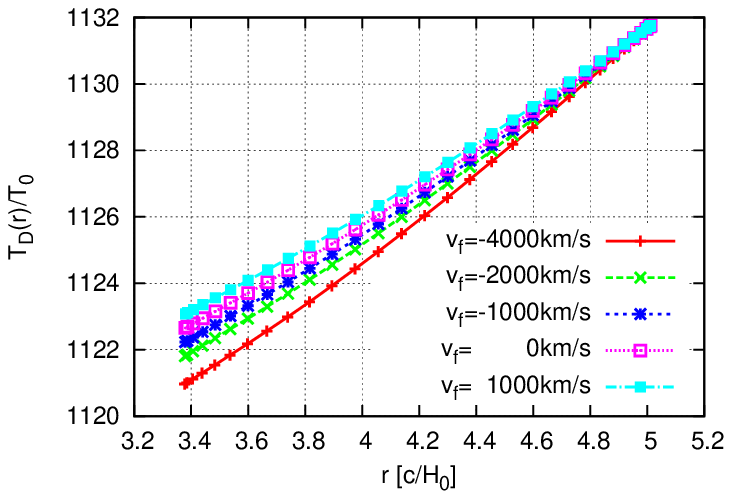}
\includegraphics[scale=1.5]{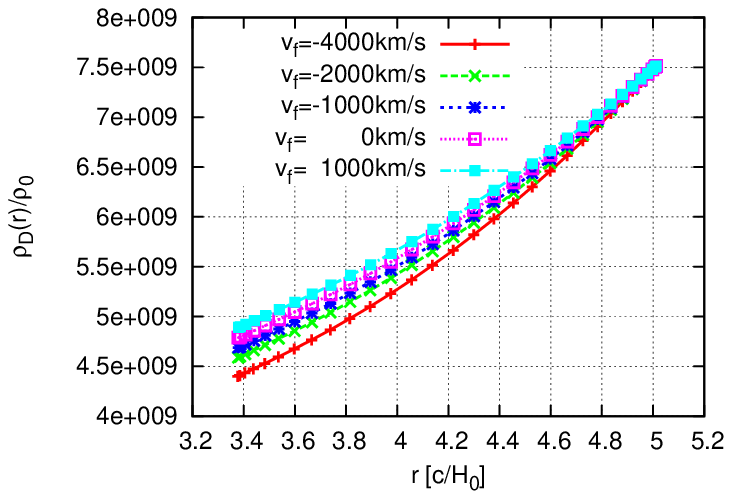}
\caption{$r$-dependence of $\Tlss(r)$ and $\rho_{\rm D}(r)$. 
}
\label{fig:alpha}
\end{center}
\end{figure}

However, it is too early to conclude from the above result
that the LTB-type anti-Copernican cosmology is consistent 
with observation. There is a very important point to be reminded.
In Fig.~\ref{fig:alpha}, we depict the temperature
and density distributions on the decoupling hypersurface,
$\Tlss(r)$ and $\rho_{\rm D}(r):=\rho(t_{\rm D}(r),r)$. 
We can see from Fig.~8 that there is a factor of two difference 
in the matter density $\rho_D(r)$ on the decoupling hypersurface
over the range of $r$ of our interest, while the temperature $T_D(r)$ is
nearly uniform. Thus the uniform temperature hypersurfaces 
are far different from uniform density hypersurfaces, implying the existence
of an isocurvature perturbation with the amplitude of order unity.
In particular, on a $\rho=$const. hypersurface, 
the distribution of the radiation is highly inhomogeneous.
Therefore it would be wrong to assume the 
evolution of the universe to be the same as in a homogeneous 
and isotropic universe in the radiation dominated era. 

The above result leads to a significant difficulty in predicting 
almost all observables related to the structure formation.
To begin with, it implies that an LTB model in which the photon
energy density is neglected ceases to be a good approximation
to the universe at the decoupling epoch.
Therefore we first have to obtain spherically symmetric background
universe models with radiation and dust, in which the radial 
profiles of radiation and dust are different.
Then we need to perform cosmological perturbation analyses
of these models in order to make theoretical predictions.
This is an issue well beyond the scope of this
paper. We just mention that there have been some advancements 
in this direction recently~\cite{Clarkson:2010uz,Clarkson:2010ej}.

\section{Summary and Discussion}
\label{sec:sudi}
In this paper, we have discussed the kSZ effect in an 
LTB universe model whose distance-redshift relation observed 
at the center agrees with the concordance $\Lambda$CDM model 
at redshift $z\lesssim2$.
Since the LTB universe model is spherically symmetric,
it is natural to assume that physical quantities at the decoupling 
epoch are functions of the radial coordinate. 
We have assumed the decoupling 
between baryons and photon occurs instantaneously, or in other words, 
it occurs on a single spacelike hypersurface. 
To determine this decoupling hypersurface
we have adopted Gamow's criterion, $H=\Gamma$, for simplicity,
where $H$ is the Hubble parameter and $\Gamma$ is the rate
of collisions between photons and electrons.
However, this criterion is not sufficient to determine 
the decoupling hypersurface. Instead, we have found that
we need two more conditions for the physical quantities 
(the energy densities of dark matter, baryons and photons, 
the temperature of photon) on the decoupling hypersurface. 

In \S\ref{sec:simplest}, 
we have assumed the adiabatic perturbation scenario,
namely we have fixed the ratios between two of 
any physical quantities (e.g., baryon-to-photon ratio)
to be constant on the decoupling hypersurface. 
Then, it turned out that the CMB dipole anisotropy 
in each cluster rest frame 
is much larger than the observational data deduced from
the kSZ effect. 
Therefore we conclude that our LTB universe model 
in the context of the conventional adiabatic perturbation
scenario is ruled out. 

Then in \S\ref{sec:fidcl}, we have introduced 
radial non-adiabatic 
(isocurvature) inhomogeneities in the non-relativistic matter 
on the decoupling hypersurface. 
We assumed that the baryon-to-photon ratio is 
constant on the decoupling hypersurface. 
Then, we have shown that appropriate spatial distributions of
isocurvature inhomogeneities can improve the consistency with 
observational results. 
It should be noted that we have introduced only a radial inhomogeneity. 
Hence it does not contribute directly to the CMB temperature anisotropy
observed at the center. 
Actually, we may have a much larger value of the likelihood  
than the result in \cite{GarciaBellido:2008gd}. 
Thus, at present, the LTB universe model has not yet been 
ruled out completely by the observation of the kSZ effect.
However, we must be cautioned by the fact that
the necessary magnitude of isocurvature inhomogeneities 
is so large that our simple LTB universe model that
neglects the photon energy density ceases to be a good 
approximation to the universe at the decoupling. 
Therefore our result should be taken not as evidence
but as an indication that the presence of large 
isocurvature perturbations may save the LTB universe model. 

Finally, we mention that the possibility to 
consider the off-center observer in LTB universe models. 
In \cite{Alnes:2006pf}, Alnes and Amarzguioui claimed that 
the observer has to be located within a radius of 
15Mpc from the center for the induced dipole 
to be less than that observed by the COBE satellite. 
Kodama et al. also reported similar constraints 
on several LTB models~\cite{Kodama:2010gr}. 
However, as shown in \S\ref{sec:fidcl}, 
it is possible to make an off-center observer 
in an LTB universe see a vanishing dipole anisotropy 
if we turn on radial isocurvature inhomogeneities at the 
decoupling epoch. It may be also interesting to investigate
the relation of the kSZ effect in LTB universe models
to the large scale coherent bulk flow reported by 
Kashlinsky et al.~\cite{Kashlinsky:2009dw,Kashlinsky:2008ut,Kashlinsky:2008us}.

\begin{acknowledgments}
C.Y. would like to thank the organizers and all the participants 
of the workshop LLTB2009 in KEK Tsukuba, Japan, for helpful comments. 
We would also like to thank all the participants of the 
Long-term Workshop on Gravity and Cosmology (GC2010: YITP-T-10-01)
for fruitful discussions. In particular, we thank
Dmitry Gorbunov and Valery Rubakov for an enlightening explanation
of the decoupling process.
C.Y. acknowledges the
Korea Ministry of Education, Science and Technology (MEST) for the support
of the Young Scientist Training Program at the Asia Pacific Center for
Theoretical Physics (APCTP). 
This work was also supported in part by 
Korea Institute for Advanced Study under the KIAS Scholar program,  
by the Grant-in-Aid for the Global COE Program 
``The Next Generation of Physics, Spun from Universality and Emergence''
from the Ministry of Education, Culture, 
Sports, Science and Technology (MEXT) of Japan, 
by Grant-in-Aid for Scientific Research (C), No.~21540276 
from the Ministry of Education, Science, Sports and Culture,
by JSPS Grant-in-Aid for Scientific Research (A) No.~21244033,
and by JSPS Grant-in-Aid for Creative Scientific Research No.~19GS0219.
\end{acknowledgments}

\providecommand{\href}[2]{#2}\begingroup\raggedright\endgroup


\begin{thebibliography}{10}

\bibitem{Zehavi:1998gz}
I.~Zehavi, A.~G. Riess, R.~P. Kirshner, and A.~Dekel, {\it {A Local Hubble
  Bubble from SNe Ia?}},  {\em Astrophys. J.} {\bf 503} (1998) 483,
  [\href{http://xxx.lanl.gov/abs/astro-ph/9802252}{{\tt astro-ph/9802252}}].

\bibitem{Tomita:1999qn}
K.~Tomita, {\it {Distances and lensing in cosmological void models}},  {\em
  Astrophys. J.} {\bf 529} (2000) 38,
  [\href{http://xxx.lanl.gov/abs/astro-ph/9906027}{{\tt astro-ph/9906027}}].

\bibitem{Tomita:2000jj}
K.~Tomita, {\it {A Local Void and the Accelerating Universe}},  {\em Mon. Not.
  Roy. Astron. Soc.} {\bf 326} (2001) 287,
  [\href{http://xxx.lanl.gov/abs/astro-ph/0011484}{{\tt astro-ph/0011484}}].

\bibitem{Tomita:2001gh}
K.~Tomita, {\it {Analyses of Type Ia Supernova Data in Cosmological Models with
  a Local Void}},  {\em Prog. Theor. Phys.} {\bf 106} (2001) 929--939,
  [\href{http://xxx.lanl.gov/abs/astro-ph/0104141}{{\tt astro-ph/0104141}}].

\bibitem{Celerier:1999hp}
M.-N. Celerier, {\it {Do we really see a cosmological constant in the
  supernovae data ?}},  {\em Astron. Astrophys.} {\bf 353} (2000) 63--71,
  [\href{http://xxx.lanl.gov/abs/astro-ph/9907206}{{\tt astro-ph/9907206}}].

\bibitem{Mustapha:1998jb}
N.~Mustapha, C.~Hellaby, and G.~F.~R. Ellis, {\it {Large scale inhomogeneity
  versus source evolution: Can we distinguish them observationally?}},  {\em
  Mon. Not. Roy. Astron. Soc.} {\bf 292} (1997) 817--830,
  [\href{http://xxx.lanl.gov/abs/gr-qc/9808079}{{\tt gr-qc/9808079}}].

\bibitem{Iguchi:2001sq}
H.~Iguchi, T.~Nakamura, and K.-i. Nakao, {\it {Is dark energy the only solution
  to the apparent acceleration of the present universe?}},  {\em Prog. Theor.
  Phys.} {\bf 108} (2002) 809--818,
  [\href{http://xxx.lanl.gov/abs/astro-ph/0112419}{{\tt astro-ph/0112419}}].

\bibitem{Yoo:2008su}
C.-M. Yoo, T.~Kai, and K.-i. Nakao, {\it {Solving Inverse Problem with
  Inhomogeneous Universe}},  {\em Prog. Theor. Phys.} {\bf 120} (2008)
  937--960, [\href{http://xxx.lanl.gov/abs/0807.0932}{{\tt arXiv:0807.0932}}].

\bibitem{Celerier:2009sv}
M.-N. Celerier, K.~Bolejko, A.~Krasinski, and C.~Hellaby, {\it {A (giant) void
  is not mandatory to explain away dark energy with a Lemaitre -- Tolman
  model}},  \href{http://xxx.lanl.gov/abs/0906.0905}{{\tt arXiv:0906.0905}}.

\bibitem{Kolb:2009hn}
E.~W. Kolb and C.~R. Lamb, {\it {Light-cone observations and cosmological
  models: implications for inhomogeneous models mimicking dark energy}},
  \href{http://xxx.lanl.gov/abs/0911.3852}{{\tt arXiv:0911.3852}}.

\bibitem{Alnes:2006pf}
H.~Alnes and M.~Amarzguioui, {\it {CMB anisotropies seen by an off-center
  observer in a spherically symmetric inhomogeneous universe}},  {\em Phys.
  Rev.} {\bf D74} (2006) 103520,
  [\href{http://xxx.lanl.gov/abs/astro-ph/0607334}{{\tt astro-ph/0607334}}].

\bibitem{Alnes:2005rw}
H.~Alnes, M.~Amarzguioui, and O.~Gron, {\it {An inhomogeneous alternative to
  dark energy?}},  {\em Phys. Rev.} {\bf D73} (2006) 083519,
  [\href{http://xxx.lanl.gov/abs/astro-ph/0512006}{{\tt astro-ph/0512006}}].

\bibitem{Alexander:2007xx}
S.~Alexander, T.~Biswas, A.~Notari, and D.~Vaid, {\it {Local Void vs Dark
  Energy: Confrontation with WMAP and Type Ia Supernovae}},  {\em JCAP} {\bf
  0909} (2009) 025, [\href{http://xxx.lanl.gov/abs/0712.0370}{{\tt
  arXiv:0712.0370}}].

\bibitem{Bolejko:2008cm}
K.~Bolejko and J.~S.~B. Wyithe, {\it {Testing the Copernican Principle via
  Cosmological Observations}},  {\em JCAP} {\bf 0902} (2009) 020,
  [\href{http://xxx.lanl.gov/abs/0807.2891}{{\tt arXiv:0807.2891}}].

\bibitem{Clarkson:1999yj}
C.~A. Clarkson and R.~Barrett, {\it {Does the Isotropy of the CMB Imply a
  Homogeneous Universe? Some Generalised EGS Theorems}},  {\em Class. Quant.
  Grav.} {\bf 16} (1999) 3781--3794,
  [\href{http://xxx.lanl.gov/abs/gr-qc/9906097}{{\tt gr-qc/9906097}}].

\bibitem{Caldwell:2007yu}
R.~R. Caldwell and A.~Stebbins, {\it {A Test of the Copernican Principle}},
  {\em Phys. Rev. Lett.} {\bf 100} (2008) 191302,
  [\href{http://xxx.lanl.gov/abs/0711.3459}{{\tt arXiv:0711.3459}}].

\bibitem{GarciaBellido:2008gd}
J.~Garcia-Bellido and T.~Haugboelle, {\it {Looking the void in the eyes - the
  kSZ effect in LTB models}},  {\em JCAP} {\bf 0809} (2008) 016,
  [\href{http://xxx.lanl.gov/abs/0807.1326}{{\tt arXiv:0807.1326}}].

\bibitem{GarciaBellido:2008nz}
J.~Garcia-Bellido and T.~Haugboelle, {\it {Confronting Lemaitre-Tolman-Bondi
  models with Observational Cosmology}},  {\em JCAP} {\bf 0804} (2008) 003,
  [\href{http://xxx.lanl.gov/abs/0802.1523}{{\tt arXiv:0802.1523}}].

\bibitem{GarciaBellido:2008yq}
J.~Garcia-Bellido and T.~Haugboelle, {\it {The radial BAO scale and Cosmic
  Shear, a new observable for Inhomogeneous Cosmologies}},  {\em JCAP} {\bf
  0909} (2009) 028, [\href{http://xxx.lanl.gov/abs/0810.4939}{{\tt
  arXiv:0810.4939}}].

\bibitem{Goodman:1995dt}
J.~Goodman, {\it {Geocentrism reexamined}},  {\em Phys. Rev.} {\bf D52} (1995)
  1821--1827, [\href{http://xxx.lanl.gov/abs/astro-ph/9506068}{{\tt
  astro-ph/9506068}}].

\bibitem{Godlowski:2004gh}
W.~Godlowski, J.~Stelmach, and M.~Szydlowski, {\it {Can the Stephani model be
  an alternative to FRW accelerating models?}},  {\em Class. Quant. Grav.} {\bf
  21} (2004) 3953--3972, [\href{http://xxx.lanl.gov/abs/astro-ph/0403534}{{\tt
  astro-ph/0403534}}].

\bibitem{Zibin:2008vj}
J.~P. Zibin, {\it {Scalar Perturbations on Lemaitre-Tolman-Bondi Spacetimes}},
  {\em ~} (2008) [\href{http://xxx.lanl.gov/abs/0804.1787}{{\tt
  arXiv:0804.1787}}].

\bibitem{Zibin:2008vk}
J.~P. Zibin, A.~Moss, and D.~Scott, {\it {Can we avoid dark energy?}},  {\em
  Phys. Rev. Lett.} {\bf 101} (2008) 251303,
  [\href{http://xxx.lanl.gov/abs/0809.3761}{{\tt arXiv:0809.3761}}].

\bibitem{Clifton:2009kx}
T.~Clifton, P.~G. Ferreira, and J.~Zuntz, {\it {What the small angle CMB really
  tells us about the curvature of the Universe}},  {\em JCAP} {\bf 0907} (2009)
  029, [\href{http://xxx.lanl.gov/abs/0902.1313}{{\tt arXiv:0902.1313}}].

\bibitem{Regis:2010iq}
M.~Regis and C.~Clarkson, {\it {Do primordial Lithium abundances imply there's
  no Dark Energy?}},  \href{http://xxx.lanl.gov/abs/1003.1043}{{\tt
  arXiv:1003.1043}}.

\bibitem{Kodama:2010gr}
H.~Kodama, K.~Saito, and A.~Ishibashi, {\it {Analytic formulae for the
  off-center CMB anisotropy in a general spherically symmetric universe}},
  \href{http://xxx.lanl.gov/abs/1004.3089}{{\tt arXiv:1004.3089}}.

\bibitem{Garfinkle:2009uf}
D.~Garfinkle, {\it {The motion of galaxy clusters in inhomogeneous
  cosmologies}},  {\em Class. Quant. Grav.} {\bf 27} (2010) 065002,
  [\href{http://xxx.lanl.gov/abs/0908.4102}{{\tt arXiv:0908.4102}}].

\bibitem{Clarkson:2010ej}
C.~Clarkson and M.~Regis, {\it {The Cosmic Microwave Background in an
  Inhomogeneous Universe}},  \href{http://xxx.lanl.gov/abs/1007.3443}{{\tt
  arXiv:1007.3443}}.

\bibitem{Moss:2010jx}
A.~Moss, J.~P. Zibin, and D.~Scott, {\it {Precision Cosmology Defeats Void
  Models for Acceleration}},  \href{http://xxx.lanl.gov/abs/1007.3725}{{\tt
  arXiv:1007.3725}}.

\bibitem{Biswas:2010xm}
T.~Biswas, A.~Notari, and W.~Valkenburg, {\it {Testing the Void against
  Cosmological data: fitting CMB, BAO, SN and H0}},
  \href{http://xxx.lanl.gov/abs/1007.3065}{{\tt arXiv:1007.3065}}.

\bibitem{Yoo:2010qy}
C.-M. Yoo, K.-i. Nakao, and M.~Sasaki, {\it {CMB observations in LTB universes:
  Part I: Matching peak positions in the CMB spectrum}},
  \href{http://xxx.lanl.gov/abs/1005.0048}{{\tt arXiv:1005.0048}}.

\bibitem{Tanimoto:2007dq}
M.~Tanimoto and Y.~Nambu, {\it {Luminosity distance-redshift relation for the
  LTB solution near the center}},  {\em Class. Quant. Grav.} {\bf 24} (2007)
  3843--3857, [\href{http://xxx.lanl.gov/abs/gr-qc/0703012}{{\tt
  gr-qc/0703012}}].

\bibitem{Weinberg:2008zzc}
S.~Weinberg, {\it {Cosmology}}, . Oxford, UK: Oxford Univ. Pr. (2008) 593 p.

\bibitem{Holzapfel:1997ui}
W.~L. Holzapfel {\em et.~al.}, {\it {Limits on the Peculiar Velocities of Two
  Distant Clusters Using the Kinematic Sunyaev-Zel'dovich Effect}},  {\em
  Astrophys. J.} {\bf 481} (1997) 35--48,
  [\href{http://xxx.lanl.gov/abs/astro-ph/9702223}{{\tt astro-ph/9702223}}].

\bibitem{Benson:2003va}
B.~A. Benson {\em et.~al.}, {\it {Peculiar Velocity Limits from Measurements of
  the Spectrum of the Sunyaev-Zel'dovich Effect in Six Clusters of Galaxies}},
  {\em Astrophys. J.} {\bf 592} (2003) 674--691,
  [\href{http://xxx.lanl.gov/abs/astro-ph/0303510}{{\tt astro-ph/0303510}}].

\bibitem{Kitayama:2003mq}
T.~Kitayama {\em et.~al.}, {\it {Exploring Cluster Physics with High-resolution
  Sunyaev- Zel'dovich Effect Images and X-ray Data: A Case of the Most X-ray
  Luminous Galaxy Cluster RXJ1347-1145}},  {\em Publ. Astron. Soc. Jap.} {\bf
  56} (2004) 17--28, [\href{http://xxx.lanl.gov/abs/astro-ph/0311574}{{\tt
  astro-ph/0311574}}].

\bibitem{Clarkson:2010uz}
C.~Clarkson and R.~Maartens, {\it {Inhomogeneity and the foundations of
  concordance cosmology}},  {\em Class. Quant. Grav.} {\bf 27} (2010) 124008,
  [\href{http://xxx.lanl.gov/abs/1005.2165}{{\tt arXiv:1005.2165}}].

\bibitem{Kashlinsky:2009dw}
A.~Kashlinsky, F.~Atrio-Barandela, H.~Ebeling, A.~Edge, and D.~Kocevski, {\it
  {A new measurement of the bulk flow of X-ray luminous clusters of galaxies}},
   {\em Astrophys. J.} {\bf 712} (2010) L81--L85,
  [\href{http://xxx.lanl.gov/abs/0910.4958}{{\tt arXiv:0910.4958}}].

\bibitem{Kashlinsky:2008ut}
A.~Kashlinsky, F.~Atrio-Barandela, D.~Kocevski, and H.~Ebeling, {\it {A
  measurement of large-scale peculiar velocities of clusters of galaxies:
  results and cosmological implications}},  {\em Astrophys. J.} {\bf 686}
  (2008) L49--L52, [\href{http://xxx.lanl.gov/abs/0809.3734}{{\tt
  arXiv:0809.3734}}].

\bibitem{Kashlinsky:2008us}
A.~Kashlinsky, F.~Atrio-Barandela, D.~Kocevski, and H.~Ebeling, {\it {A
  measurement of large-scale peculiar velocities of clusters of galaxies:
  technical details}},  {\em Astrophys. J.} {\bf 691} (2009) 1479--1493,
  [\href{http://xxx.lanl.gov/abs/0809.3733}{{\tt arXiv:0809.3733}}].

\end{thebibliography}

\end{document}